\documentclass{iopjournal}
\usepackage[german,american,english]{babel}
\usepackage{graphicx}
\usepackage{graphics}
\usepackage{dcolumn}
\usepackage{multirow}
\usepackage{bm}
\usepackage{latexsym}
\usepackage{amssymb}
\usepackage{amsmath}
\usepackage{amsfonts}
\usepackage{layout}
\usepackage{verbatim}
\usepackage{epsfig}
\usepackage{graphicx}
\usepackage{amsbsy}
\usepackage{xcolor}
\usepackage[caption=false]{subfig}
\newcommand{\bea}{\begin{eqnarray*}}
	\newcommand{\eea}{\end{eqnarray*}}
\newcommand{\bne}{\begin{equation*}}
\newcommand{\ede}{\end{equation*}}

\newcommand{\bnen}{\begin{equation}}
\newcommand{\eden}{\end{equation}}
\newcommand{\bean}{\begin{eqnarray}}
\newcommand{\eean}{\end{eqnarray}}
\newcommand{\bsen}{\begin{subequations}}
	\newcommand{\esen}{\end{subequations}}

\newcommand{\ba}{\arraycolsep 0.3ex \begin{array}{rl}}
\newcommand{\ea}{\end{array}}

\newcommand{\bna}{\begin{array}}
	\newcommand{\eda}{\end{array}}
\newcommand{\bnm}{\begin{enumerate}}
	\newcommand{\edm}{\end{enumerate}}

\newcommand {\bkt} [1] {\langle #1 \rangle}

\def\frac#1#2{{{#1 \over #2}}}

\def\frac#1#2{{\textstyle{#1 \over #2}}}

\def\Tra{\mathop{\textsf{Tr}}}

\def\Tra{\mathop{\textsf{Tr}}}

\def\bmapright#1{\ \smash{\mathop{\hbox to 25pt{\rightarrowfill}}\limits_{#1}}\ }

\begin{document}

\articletype{Letter}

\title{Theory of extrinsic contributions to the full orbital current}

\author{James H Cullen$^{1,*}$\orcid{0000-0003-2297-3471} and Dimitrie Culcer$^{1}$\orcid{0000-0002-2342-0396}}

\affil{$^1$School of Physics, The University of New South Wales, Sydney 2052, Australia}

\affil{$^*$Author to whom any correspondence should be addressed.}

\email{james.cullen@unsw.edu.au}


\begin{abstract}
The orbital Hall effect (OHE) underpins the emerging field of orbitronics. A complete quantum-mechanical evaluation of the usual orbital-current operator must retain all matrix elements of the position operator, including its band-diagonal differential part; intrinsic calculations based on this full evaluation can differ by orders of magnitude from the conventional truncation. A relaxation-time estimate of disorder effects on the full current has been reported, but a microscopic treatment of extrinsic scattering has remained unavailable because the position operator requires wavevector-off-diagonal density-matrix elements. Here we develop such a theory by solving the quantum kinetic equation for the wavevector-off-diagonal density matrix, including band-structure, field-corrected side-jump, and skew-scattering terms within the non-crossing approximation. We apply the theory to a massive Dirac cone and to the same model with a particle-hole-symmetry-breaking quadratic term. Disorder-generated contributions of order $\tau^0$ are generically comparable to the intrinsic current in the metallic regime and cannot be separated from it by simple disorder-strength scaling. For the bare massive Dirac cone the surviving non-crossing extrinsic correction cancels the conventional intrinsic value and leaves a total current exactly twice the quantum correction, whereas particle-hole asymmetry activates conventional skew scattering for disorder with a non-zero third moment, which dominates in sufficiently clean samples. The wavevector-off-diagonal construction is independent of the two-band model and provides a general route to disorder corrections for position-dependent observables in multiband solids.
\end{abstract}

\section{Introduction}

The generation and transport of electron orbital angular momentum (OAM) have become major topics of research with the advent of orbitronics \cite{Orbitronics-in-action, Rhonald-Rev, OC-Rev-EL-2021-Yuriy, wang2024orbitronics}. The technological driver of this field is the orbital torque, in which a nonequilibrium OAM density or current exerts a torque on an adjacent magnetic material \cite{OT-PRR-2020-Hyun-Woo}, in close analogy with the spin-torque mechanisms of spintronic devices \cite{CI-SOT-RMP-2019-Manchon, brataas2012current, gambardella2011current, kohno2006microscopic, tatara2007spin}. The primary avenue for realising this torque is the orbital Hall effect (OHE), a transverse current of OAM generated by an electric field \cite{Orbitronics-PRL-2005-Shoucheng, ISHE-IOHE-PRB-2008-Inoue, OHE-PRL-2009-Inoue}. Orbital torques have since been demonstrated in a wide range of systems \cite{OT-OEE-NatComm-2018-Haibo, Exp-OT-PRR-2020, OT-FM-PRB-2021-YoshiChika, OT-NatComm-2021-Kyung-Jin, Exp-OT-CommP-2021-Byong-Guk, OHE-OT-large, L-S-OT-2023}, aided by efficient orbital-to-spin conversion at interfaces \cite{OOS-Cvert-2020-PRL-Mathias, OHE-Hetero-PRR-2022-Pietro, OHE-Binghai, ko2026magneto}, and the OHE itself has been observed directly in light metals and semiconductors \cite{Exp-OHE-Ti-Nat-2023-Hyun-Woo, PhysRevLett.131.156702, PhysRevLett.131.156703, 10.1063/5.0106988, PhysRevB.106.184406, el2023observation, santos2024negative, Inverse-OHE-weak-SOC}. More recently, attention has turned to the length and time scales governing orbital transport, relaxation and diffusion \cite{go2023long, urazhdin2026ultrashort, guan2026evidences, kang2026orbital, park2026spatiotemporally, Niels_Orbitalsplitter}, and to harnessing orbital currents for energy-efficient magnetization switching and memory applications \cite{gupta2025harnessing, jamshed2026energy, zhang2026harnessing, peng2025unconventional, wang2025orbitalpumping}. On the theoretical side, intrinsic calculations of the OHE have been performed for a wide range of model band structures \cite{ISOHE-PRL-2018-Hyun-Woo, IOHE-Metal-PRB-2018-Hyun-Woo, IOHE-PRB-2021-Giovanni, OH-phase-TMD-PRB-2020-Tatiana, canonico2020two, OHE-BiTMD-PRL-2021-Tatiana, BiTMD-OHE-PRB-2022-Giovanni&Tatiana, OHE-PRB-2022-Manchon}, together with first-principles, materials-specific predictions \cite{OHE-metal-PRM-2022-Oppeneer, cullen2025giant, lee2025universal, sun2024theory, cysne2026orbital}. In contrast, the role of disorder in the OHE has only begun to be addressed \cite{Hong-OHE-PRL, veneri2024extrinsic, RS-OHE-disorder, Tangping, PhysRevB.108.245105}.

Theoretical calculations of the orbital Hall current require the nonequilibrium expectation value of the orbital-current operator in response to an electric field \cite{Orbitronics-PRL-2005-Shoucheng,ISHE-IOHE-PRB-2008-Inoue,OHE-PRL-2009-Inoue}. We use the microscopic OAM operator $\bm L=\frac{1}{2}(\bm r\times\bm v-\bm v\times\bm r)$ and the usual symmetrised orbital-current operator $j^a_b=\frac{1}{2}\{L^a,v^b\}$. It is useful to distinguish three approaches to evaluating this operator in solids: the atom-centred approximation (ACA), the conventional Bloch-space truncation, and the full Bloch-space evaluation. The ACA retains only local OAM within atom-centred regions \cite{lee2026anatomy}; while it can be accurate when the OAM is strongly localised, its validity worsens for more complicated unit cells and when itinerant circulation is important \cite{lee2026anatomy, OM-Berry-PRB-2016-Mokrousov}. Its practical advantage is that it is readily implemented in realistic DFT calculations. The conventional Bloch-space approach uses OAM matrix elements extracted from the equilibrium OAM expectation value \cite{Hong-OHE-PRL, OHE-BiTMD-PRL-2021-Tatiana, BiTMD-OHE-PRB-2022-Giovanni&Tatiana, IOHE-PRB-2021-Giovanni, OHE-PRB-2022-Manchon, PhysRevB.108.075427-Manchon, PhysRevLett.130.116204, OHE-Topology-2023}. This procedure retains the interband Berry-connection part of the position operator but omits band-diagonal position matrix elements, which are differential operators in the Bloch representation. Ref.~\cite{liu2025quantumOHE} showed that a direct evaluation of the same orbital-current operator with the full position operator produces additional ``quantum corrections'' that dominate the conventional contribution in several models \cite{liu2025quantumOHE, cullen2025giant, cullen2026orbital}. Thus the distinction is not between two different microscopic current operators, but between a truncated and a complete evaluation of the usual operator. The non-conservation of OAM is a separate issue: the construction of a conserved orbital current analogous to the proper spin current remains an open problem and lies beyond the scope of this work.

A complete description of the orbital Hall effect must include extrinsic mechanisms due to impurity scattering. Disorder can generate Hall-response terms of order $\tau^0$, the same disorder scaling as the intrinsic contribution, as is well established for the spin Hall \cite{inoue2004suppression, mishchenko2004spin, raimondi2005, liu2006vanishing, sugimoto2006, hankiewicz2006} and anomalous Hall \cite{Inoue-AHE-PhysRevLett.97.046604, sinitsyn2007, sinitsyn2007semiclassical, nagaosa2010anomalous} effects. A relaxation-time estimate of extrinsic corrections to the full orbital current was given in Ref.~\cite{cullen2025giant}, while microscopic disorder corrections have been studied for the conventional orbital current \cite{Hong-OHE-PRL,veneri2024extrinsic}. What has been missing is a microscopic disorder theory for the full current in which the band-diagonal position operator is retained. The central technical difficulty is that this sector requires the density matrix at infinitesimally different wavevectors, $\rho_{\bm k_+\bm k_-}$, so the collision integral itself must be formulated and expanded away from the wavevector diagonal.

In this work we develop a microscopic theory of extrinsic contributions to the full internal orbital current. We solve the quantum kinetic equation for wavevector-off-diagonal density-matrix elements, following a strategy introduced for the proper spin current \cite{PhysRevB.108.245418}, and retain the leading band-structure, field-corrected side-jump, and skew-scattering terms within the non-crossing approximation. We apply the theory to a massive Dirac cone and to the same model with a particle-hole-symmetry-breaking quadratic correction. Two central findings emerge. First, disorder-generated terms of order $\tau^0$ are generically comparable to the intrinsic response in the metallic regime and therefore cannot be separated from it by simple scaling with the disorder strength. Second, the extrinsic mechanisms differ qualitatively from those obtained with the conventional truncation. For the bare massive Dirac cone, the skew-scattering and field-corrected side-jump terms vanish and the surviving non-crossing Born contribution gives $j_L=2\Delta j_{1}$. Once particle-hole symmetry is broken, skew-scattering and side-jump contributions become finite, and conventional skew scattering can dominate in the clean limit when the disorder has a non-zero third moment. These results show that quantitative orbital-Hall predictions require disorder and the complete position operator to be treated on the same footing, and establish a framework that can be applied to general multiband Hamiltonians.

\section{The orbital current operator and effective displacement}

The single-particle density matrix $\rho$ encodes both the occupation of Bloch states and the coherence between them. The expectation value of an operator $\hat O$ is determined by $\Tra[\hat O\rho]$, so the physical content of a linear response can often be read directly from the part of $\rho$ that it probes. We calculate the current of the internal Bloch-electron OAM. To make the origin of this observable explicit, we separate the usual microscopic OAM into internal and centre-of-mass parts,
\begin{equation}
    \bm L=\bm L_{\rm int}+\bm L_{\rm CM},\qquad
    \bm L_{\rm int}=\frac{1}{2}\left[(\bm r-\bm R_{\rm CM})\times\bm v-\bm v\times(\bm r-\bm R_{\rm CM})\right],\qquad
    \bm L_{\rm CM}=\bm R_{\rm CM}\times\bm v\,.
\end{equation}
The internal orbital current is $j^\alpha_\delta=\frac{1}{2}\{L_{{\rm int},\alpha},v_\delta\}$. Thus rigid translation of the whole wave packet is excluded, whereas the position operator entering $\bm L_{\rm int}$ is still evaluated with all its band-diagonal and band-off-diagonal matrix elements. As shown in Ref.~\cite{liu2025quantumOHE}, its expectation value can be written in terms of the internal effective displacement as
\begin{equation}\label{Eq:OHcurrent}
    \bkt{j^a_\delta} = \frac{\epsilon_{abc}}{4} \, {\rm tr} \, \int \frac{d^dk}{(2\pi)^{d}} \, \bigg( 2\{ v_\delta, v_c\} \Xi^{\rm int}_b + i \, \bigg[v_c,  \frac{Dv_\delta}{Dk_b} \bigg] \rho \bigg),
\end{equation}
where $\rho$ is the single-particle density matrix. We first evaluate the complete effective displacement $\bm \Xi$ \cite{cullen2025quantum},
\begin{equation}\label{Eq:Xi}
    \bm \Xi_{\bm k} = i\frac{\partial \rho_{\bm k_+ \bm k_-}}{\partial \bm Q}\Big|_{\bm Q \rightarrow 0} + \frac{1}{2}\{\boldsymbol{\mathcal{R}}_{\bm k},\rho_{\bm k}\}\,,
\end{equation}
where $\bm k_\pm=\bm k\pm\bm Q/2$ and $\mathcal{R}$ is the Berry connection with matrix elements $\mathcal{R}^{i,mn}_{\bm k} = i\langle u_{m \bm k}|\partial u_{n\bm k}/\partial k^i\rangle$. The internal displacement in Eq.~(\ref{Eq:OHcurrent}) is $\bm\Xi^{\rm int}=\bm\Xi-\bm\Xi^{\rm CM}$, where $\bm\Xi^{\rm CM}$ is the explicitly identifiable free-flight displacement of the packet centre discussed below. The physical content of Eq.~(\ref{Eq:Xi}) is simple. The wavevector-diagonal density matrix $\rho_{\bm k}$ determines the band occupation as well as inter-band mixing at each wavevector, but it does not by itself contain all position information. The missing information comes from the phase and amplitude structure connecting $\bm k+\bm Q/2$ to $\bm k-\bm Q/2$. Generally, the first part of (\ref{Eq:Xi}) is associated with the itinerant displacement, while the second part with the Berry connection carries the local intracell displacement \cite{cullen2025quantum}. The conventional truncation retains only the part that survives when the position operator is replaced by its interband Berry-connection matrix elements. The full evaluation instead keeps both terms in Eq.~(\ref{Eq:Xi}); this is the structural origin of the additional quantum contributions found in Ref.~\cite{liu2025quantumOHE}.

The combination in Eq.~(\ref{Eq:Xi}) is gauge covariant. Under a wavevector-dependent change of Bloch basis $|u_{\bm k}\rangle\rightarrow |u_{\bm k}\rangle V_{\bm k}$, the derivative term acquires $-\frac{1}{2}\{iV_{\bm k}^\dagger\partial_{\bm k}V_{\bm k},\rho_{\bm k}\}$, while the Berry-connection term acquires the opposite contribution. Hence $\bm\Xi_{\bm k}\rightarrow V_{\bm k}^\dagger\bm\Xi_{\bm k}V_{\bm k}$. The expression involving the centre of mass is separately invariant in the non-degenerate notation used here; degenerate subspaces are treated by replacing individual-band components with their spectral projectors. A fuller demonstration is given in the Supplemental Material.

\section{Method}

To evaluate the effective displacement, the nonequilibrium density matrix is required at infinitesimally different wavevectors. Operationally, we introduce a weak finite-wavevector electric perturbation, factor out its momentum-conserving distribution, and work with the resulting reduced Wigner density matrix $\rho_{E_i,\bm k_+\bm k_-}$. The derivative with respect to $\bm Q$ is taken before the homogeneous limit $\bm Q\rightarrow0$; no derivative of a translational $\delta(\bm Q)$ is treated as an ordinary function. We therefore solve the following kinetic equation to zeroth and linear order in $\bm Q$ \cite{PhysRevB.108.245418, liu2025quantumOHE}
\begin{equation} \label{eq:QKEQ}
    \frac{\partial \rho_{E_i,\bm k_+ \bm k_-}}{\partial t} +\frac{i}{\hbar}[H_0,\rho_{E_i}]_{\bm k_+ \bm k_-} + J(\rho_{E_i})_{\bm k_+ \bm k_-}+J_3(\rho_{E_i})_{\bm k_+ \bm k_-}+J_{E_i}(\rho_0)_{\bm k_+ \bm k_-} = \frac{e E_i}{\hbar} \frac{D \rho_{0,\bm k_+ \bm k_-}}{D k_i}\,.
\end{equation}
Here $H_0$ is the band Hamiltonian, $J$ is the second-order Born collision term, $J_3$ is the third-order collision term responsible for conventional skew scattering, and $J_{E_i}$ is the electric-field-corrected collision term. We take $e>0$ and $H_E=+e\bm E\cdot\hat{\bm r}$, so the electron charge is $-e$; all currents are quoted in this direct-response convention. All quantities in Eq.~(\ref{eq:QKEQ}) are disorder averaged. We assume dilute, identical, short-range scalar impurities, so the second and third disorder moments are proportional to $n_iU_0^2$ and $n_iU_0^3$, respectively. The conventional skew term therefore requires a non-zero third disorder moment and is absent for Gaussian disorder. We expand the equation to linear order in $\bm Q$ and retain the collision terms required within the non-crossing approximation. 

The matrix elements of the scattering term $J$ evaluated to the second order Born approximation are \cite{Interband-Coherence-PRB-2017-Dimi}
\begin{equation}\label{eq:Born}
    J(\rho)^{mm^\prime}_{\bm k_+\bm k_-} = \frac{1}{\hbar^2}\int_0^\infty dt^\prime \langle[U,[e^{-iH_0t^\prime/\hbar}Ue^{iH_0t^\prime/\hbar},\langle\rho\rangle]]\rangle^{mm^\prime}_{\bm k_+ \bm k_-}\,,
\end{equation}
where $U$ is the disorder potential and $\langle\cdots\rangle$ indicates the disorder average. We derive Eq.~(\ref{eq:Born}) and the corresponding third-order and field-corrected collision terms in the Supplemental Material. At $\bm Q=0$ our field-corrected term reduces to the collision integral derived in Ref.~\cite{JE-PRR-Rhonald-2022}; its linear-in-$\bm Q$ extension is required here for the internal displacement. The collision terms and the density matrix are expanded to linear order in $\bm Q$, which supplies the information required for the effective displacement. 

We work in the weak-scattering limit and use non-degenerate band notation, decomposing the density matrix as $\rho=n+S$, where $n$ is band diagonal and $S$ is band off diagonal. For degenerate bands the same construction is obtained by defining $n$ and $S$ as the projector-diagonal and projector-off-diagonal parts with respect to the distinct energy subspaces, and using covariant derivatives within each subspace \cite{ma2024spin}. The leading $n_{E_i}$ is the Fermi-surface Drude response, while the leading $S_{E_i}$ contains the intrinsic interband-coherence response. Subleading Born terms, the field-corrected collision integral $J_{E_i}$, and the third-order term $J_3$ generate side-jump- and skew-scattering-type corrections. In diagrammatic language $J_{E_i}$ corresponds to electric-field insertions in the impurity vertex, while $J_3$ generates the non-crossing Y diagrams. We do not include the crossing X and $\Psi$ diagrams \cite{veneri2024extrinsic, wang2026inter}; these can also contribute at order $\tau^0$, so exact numerical identities below refer to the retained non-crossing terms. The wavevector-off-diagonal construction and the inability to identify all $\tau^0$ terms as intrinsic do not rely on this omission.

The band-diagonal $\bm Q$-linear hierarchy contains a conserved density mode, so its free-streaming part cannot be assigned a finite homogeneous steady-state displacement by simply inverting the collision operator. This does \emph{not} mean that the corresponding terms should be deleted from the density matrix. We retain the complete density matrix in Eq.~(\ref{eq:QKEQ}), including the free-streaming pieces, because they enter subsequent scattering processes. The separation follows instead from the definition of $\bm L_{\rm int}$: when $\bm\Xi$ is used as its position lever arm, the projection onto the conserved free-flight mode is $\bm\Xi^{\rm CM}$ and does not enter $\bm\Xi^{\rm int}$.

Two contributions are selected by this criterion as centre-of-mass free flight. The first is the $\tau^2$ term whose $\bm Q$-linear streaming source is $-i\bm Q\cdot\bm v_d\,n^{(1)}_{\bm E,0}$: the Drude distribution, already of order $\tau$, streams for another collision time and produces an additional displacement of the wave-packet centre. The same physical structure occurs one order lower, in the $\tau$ term driven by $-i\bm Q\cdot\bm v_d\,n^{(0)}_{\bm E,0}$. These are precisely the two contributions assigned to $\bm\Xi^{\rm CM}$. Nothing is removed from $n_{\bm E}$ or $S_{\bm E}$ themselves, and no other diagonal contribution is assigned to $\bm\Xi^{\rm CM}$ unless it can be shown independently to represent the same rigid translation. Local terms contained in $\{\boldsymbol{\mathcal R},\rho\}$ and scattering-induced relative displacements are retained in $\bm\Xi^{\rm int}$. Under a shift of origin both $\bm\Xi$ and the centre coordinate shift by the same amount, so their difference and the internal current are origin independent. The transient solution in the Supplemental Material verifies directly that the quantity that has been selected is the unbounded drift of the carrier centre of mass.

\section{Results}
We now apply the formalism to two closely related Dirac models. The extrinsic corrections considered below are Fermi-surface contributions.

\begin{figure}[t]
\begin{center}
\includegraphics[trim=0cm 0cm 0cm 0cm, clip, width=0.9\columnwidth]{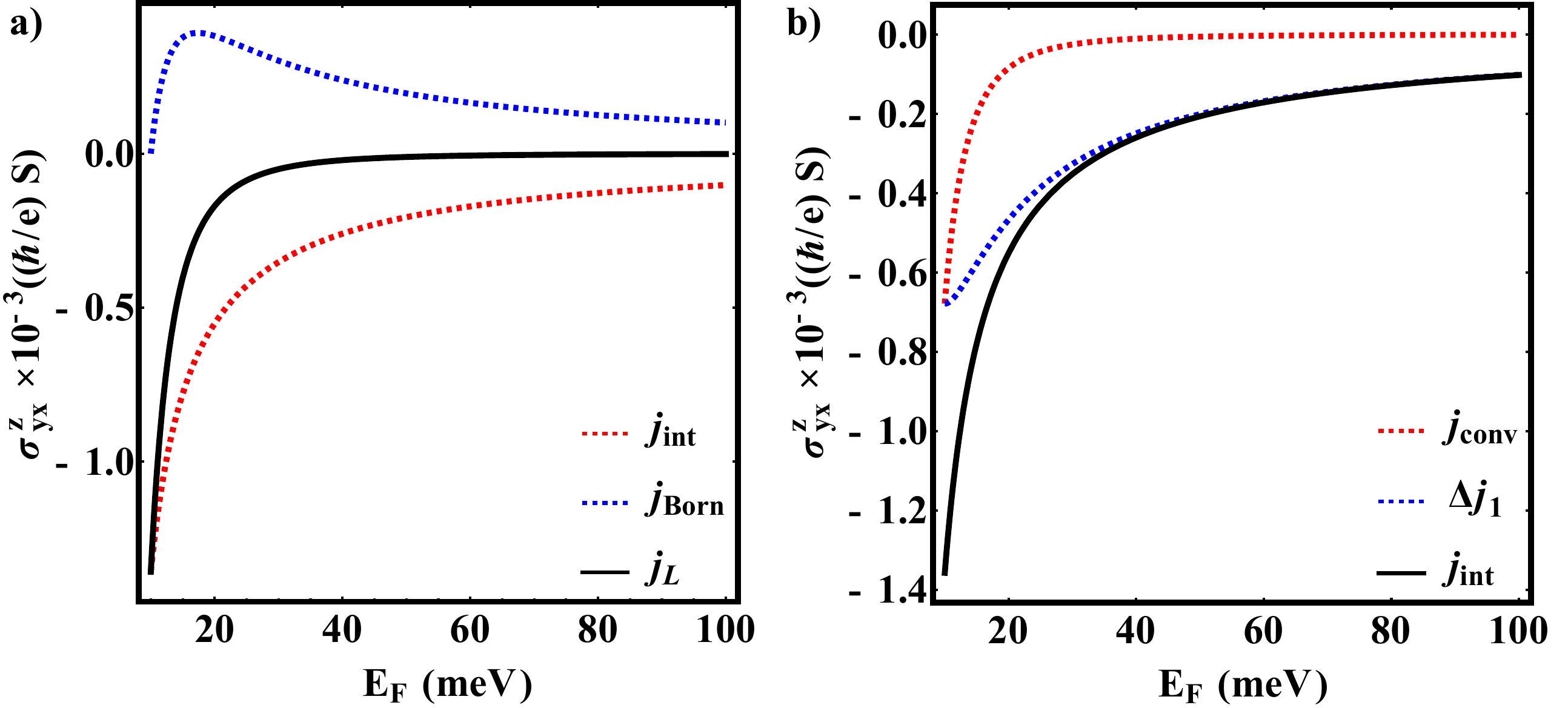}
\caption{\label{MDC_OHE}
The orbital conductivity vs the Fermi energy for the massive Dirac cone model. We plot a) the intrinsic, extrinsic and total orbital conductivity and b) the intrinsic conductivity broken down into the conventional term and the first quantum correction. The horizontal axis is the Fermi energy measured from the conduction-band edge; in Eqs.~(\ref{Eq:intJ_MDC})--(\ref{Eq:totalJ_MDC}) $E_F$ is measured from the Dirac point. Here we use $\alpha=4$ eV\AA and $m=10$ meV.}
\end{center}
\end{figure}

\subsection{Model 1: Massive Dirac cone}
The massive Dirac cone model we consider is
\begin{equation}
    H_0=\alpha(k_y\sigma_x-k_x \sigma_y)+m\sigma_z\,.
\end{equation}
where $\sigma_i$ are the Pauli matrices. The first observation is that the second term in Eq.~(\ref{Eq:OHcurrent}) vanishes for this model. The covariant derivative of the velocity is equivalent to the second covariant derivative of the Hamiltonian, and the Hamiltonian is linear in $\bm k$. The current therefore reduces to 
\begin{equation}
    \frac{\epsilon_{abc}}{2} \, {\rm tr} \, \int \frac{d^dk}{(2\pi)^{d}} \,   \{ v_\delta,  v_c\} \Xi^{\rm int}_b \,.
\end{equation}
This observation already shows that the full calculation will differ substantially from previous conventional approaches to the orbital current in Refs.~\cite{Hong-OHE-PRL,veneri2024extrinsic}. With the conventional truncation, the orbital-current expectation value retains a $\text{Tr } \epsilon_{abc}/4[v_c ,[\mathcal{R}_b,v_\delta]]\rho$ from the covariant derivative. This term captures the majority of the skew-scattering and side-jump contributions previously reported.

The velocity anticommutators are either zero or angle independent and proportional to the identity. After the centre-of-mass component is separated from the lever arm, the relevant contribution to $\bm\Xi^{\rm int}$ is generated by the off-diagonal density matrix at $\bm Q=0$ through $\{\boldsymbol{\mathcal R},\rho\}$. The third-order term $J_3$ and the field-corrected collision term $J_{E_i}$ do not contribute to the orbital current in this model. The only surviving extrinsic correction comes from $J(n_{E_i})^{od}$, which is the band-structure side-jump contribution. The analytical expressions for each contribution to the orbital current are
\begin{align}
    \label{Eq:intJ_MDC}\bkt{j^z_{y,int}}=&-\frac{e E_x \alpha^2 (m^2+3E_F^2)}{48 \pi \hbar^2 E_F^3}\,,\\
    \label{Eq:extJ_MDC}\bkt{j^z_{y,Born}}=&\frac{e E_x \alpha^2 (m^2 - E_F^2)}{16 \pi \hbar^2 E_F^3}\,,\\
    \label{Eq:totalJ_MDC}\bkt{j^z_{y,L}}=&-\frac{e E_x \alpha^2 (3E_F^2 - m^2)}{24 \pi \hbar^2 E_F^3}\,.
\end{align}
Our results for the orbital conductivity in the massive Dirac cone model are plotted in Fig.~\ref{MDC_OHE}. As shown in the figure, the extrinsic correction is small near the band edge, vanishes at the conduction-band edge, and is absent when the Fermi energy lies in the gap. This is to be expected as the extrinsic contribution is a Fermi surface effect. For Fermi energies away from the gap the extrinsic correction carries the same sign as the intrinsic contribution and doubles the quantum correction, as Eq.~(\ref{Eq:totalJ_MDC}) shows, the total current scales as $E_F^{-1}$. Figure~\ref{MDC_OHE} shows the conduction-band result; the valence-band response is obtained by the corresponding particle-hole reflection.

Notably, for this model the intrinsic current can be written as $j_{int}=j_{conv}+\Delta j_1$, where $\Delta j_1$ is the quantum correction plotted in Fig.~\ref{MDC_OHE}b). The retained Born contribution satisfies $j_{Born}=\Delta j_1-j_{conv}$. Hence the conventional current cancels between the intrinsic and Born terms while the quantum correction doubles. The first quantum correction to the orbital current is
\begin{equation}
    \bkt{\Delta j^z_{y,1}}=-\frac{e E_x \alpha^2 (3E_F^2 - m^2)}{48 \pi \hbar^2 E_F^3}\,,
\end{equation}
this is exactly half the total non-crossing current. As such, the retained band-structure side-jump correction gives exactly $j_L=2\Delta j_1$.

\subsection{Model 2: Massive Dirac cone with PH-symmetry breaking term}

\begin{figure}[t]
\begin{center}
\includegraphics[trim=0cm 0cm 0cm 0cm, clip, width=0.9\columnwidth]{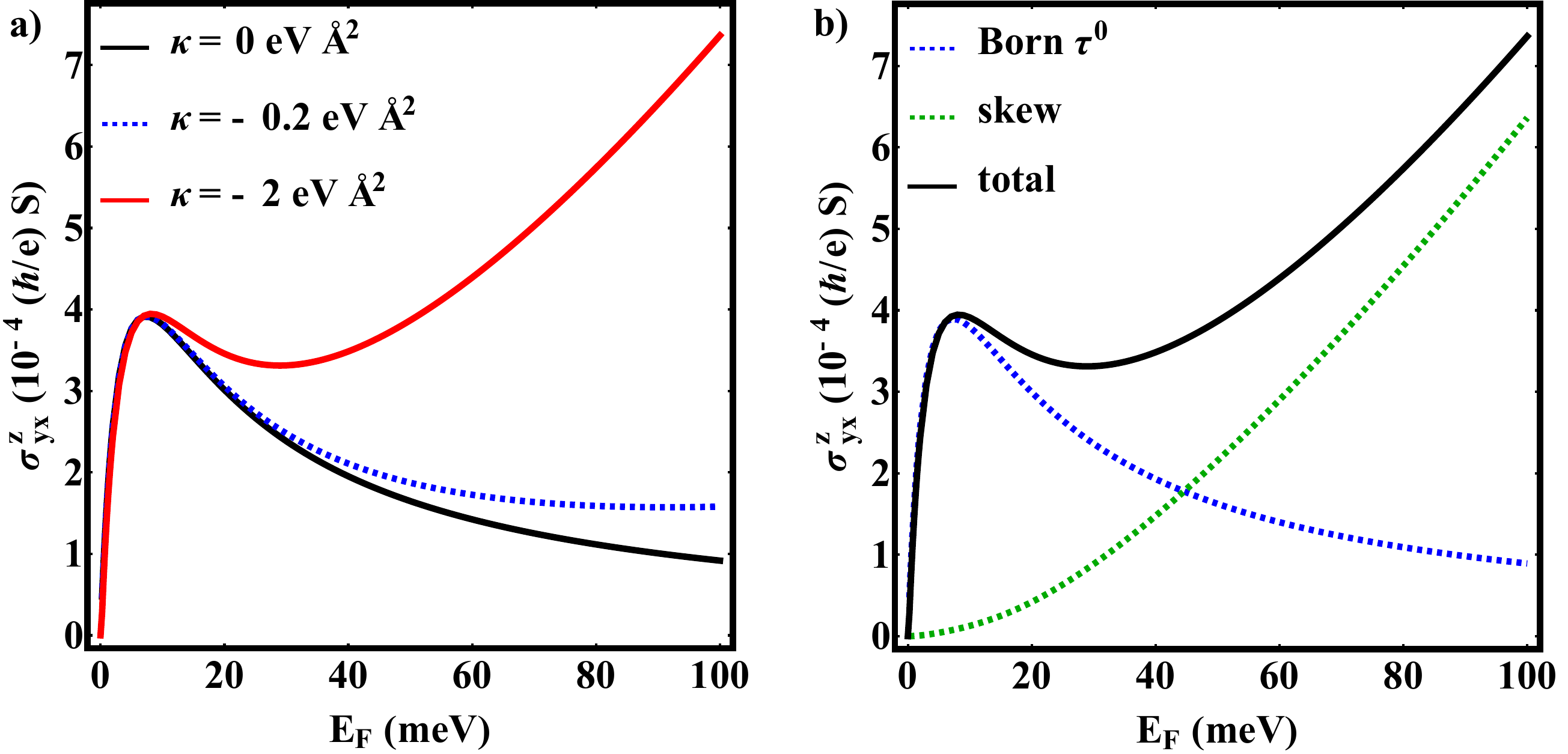}
\caption{\label{Total_ext_TMDCPH}
The extrinsic orbital conductivity vs the Fermi energy for the massive Dirac cone model with the particle-hole symmetry breaking term $H_{PH}$. We plot the extrinsic orbital conductivity a) for different values of $\kappa$ and b) its separate contributions for $\kappa=-2$ eV\AA$^2$ . Here we use $\alpha=4$ eV\AA, $m=10$ meV, $\tau=1$ ps and $n_i=10^{11}$ cm$^{-2}$.}
\end{center}
\end{figure}

In this section, we include a quadratic term in the massive Dirac cone Hamiltonian that breaks particle-hole symmetry;
\begin{equation}
    H_{PH}=\kappa k^2\,\sigma_0\,.
\end{equation}
The third-order and field-corrected collision terms do not contribute to the orbital current of the bare massive Dirac cone, but become finite after the quadratic $H_{PH}$ term is added. Real materials with Dirac cone-like band structures near the Fermi surface will have quadratic corrections; this model shows that skew and side-jump contributions need not vanish once particle-hole symmetry is broken. We take $H_{PH}$ to be a small correction and restrict the parameters so that the bands do not overlap; interband impurity scattering is then absent, allowing us to isolate the physics associated with the Dirac point. The on-shell delta functions reduce the numerical calculation to the single, non-overlapping Fermi contour, and all extrinsic curves shown here are Fermi-surface contributions.

The extrinsic orbital currents are plotted in Fig.~\ref{Total_ext_TMDCPH}; we find that there are significant contributions from skew-scattering as well as the Born scattering terms. For this model the field-corrected side-jump term is negligible: for the parameters of Fig.~\ref{Total_ext_TMDCPH} it is of order $10^{-6}$–$10^{-5}\,(\hbar/e)$ S, and we therefore retain it in all totals but do not plot it separately. The skew-scattering and side-jump contributions that were absent in the bare massive Dirac cone are all linear in the parameter $\kappa$, as seen in panel a) of Fig.~\ref{Total_ext_TMDCPH}; this means that the sign of the current from these contributions will depend on the sign of $\kappa$. Because skew scattering is odd in the impurity potential, its sign also reverses under $U_0\rightarrow-U_0$; throughout the numerical calculations we take $U_0>0$.

\begin{figure}[t]
\begin{center}
\includegraphics[trim=0cm 0cm 0cm 0cm, clip, width=0.9\columnwidth]{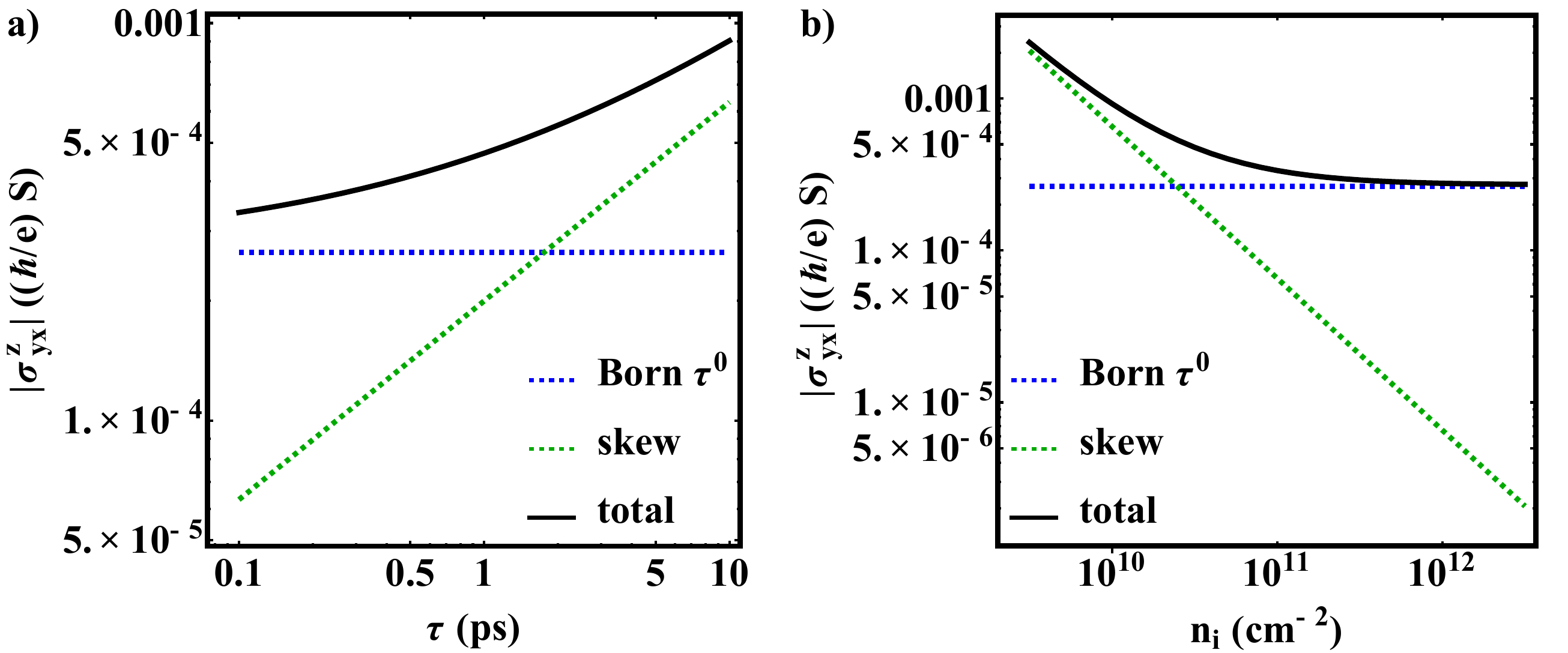}
\caption{\label{Scale_ext_TMDCPH}
The extrinsic orbital conductivity and its components vs a) the relaxation time obtained by varying $U_0$ with the impurity density fixed at $10^{11}$ cm$^{-2}$ and b) the impurity density with the disorder potential strength fixed at $U_0=200$ eV\AA$^2$, for the massive Dirac cone model with the particle-hole symmetry breaking term $H_{PH}$. Panel b) exhibits the perturbative skew-scattering scaling $j_{\rm skew}\propto\tau\propto 1/n_i$ at fixed $U_0$; in panel a) $U_0$ and $\tau$ vary together. Here we use $E_F=25$ meV, $\alpha=4$ eV\AA, $m=10$ meV and $\kappa=-2$ eV\AA$^2$.}
\end{center}
\end{figure}

Fig.~\ref{Scale_ext_TMDCPH} shows the disorder scaling of the extrinsic orbital current. The perturbative counting is most transparent with the impurity potential held fixed. Since $J_3\propto n_iU_0^3$ and each inversion of the Born collision operator contributes a factor $\tau\propto (n_iU_0^2)^{-1}$, the leading skew contribution has the structure $j_{\rm skew}\propto U_0\tau$. It is therefore of order $\tau$ in the weak-disorder expansion and, at fixed $U_0$, scales as $1/n_i$. The skew contribution is odd under $U_0\rightarrow-U_0$. Panel b) displays this $1/n_i$, equivalently $\tau$-linear, scaling. In panel a), by contrast, $n_i$ is held fixed while $U_0$ is varied and the horizontal axis is expressed in terms of the corresponding relaxation time. Since $U_0$ and $\tau$ are then varied simultaneously, panel a) should not be interpreted as defining an independent power-law dependence on $\tau$. The relaxation time $\tau$ used in this model is defined as 
\begin{equation}
    1/\tau=\frac{n_i U_0^2}{\hbar}\,\frac{2m^2+\alpha^2 k^2}{2\alpha^2(m^2+\alpha^2 k^2)^{1/2}}
\end{equation} 
\noindent where $n_i$ is the impurity density. This rate is determined by the out-scattering part of Eq.~(\ref{eq:Born}). We have excluded $\kappa$ from the scattering integrals, where they appear in the energy delta functions, for simplicity. Our numerical estimates indicate that excluding the $\kappa$ terms from these delta functions changes the current magnitude by $\lesssim 6\%$ for the parameter values considered in this work. The $\kappa$-dependence of the orbital current in this model enters through the expressions for the velocity operators and the electric potential.

Fig.~\ref{Scale_ext_TMDCPH} shows that, similar to the anomalous Hall effect, skew-scattering dominates in the dilute/clean limit \cite{nagaosa2010anomalous}. In Figs.~\ref{Total_ext_TMDCPH} and \ref{Scale_ext_TMDCPH} the impurity potentials implied by $\tau = 1$ ps range from $U_0 \approx 140-310$ eV \AA$^2$ over the Fermi energies shown, so the disorder expansion is well controlled. Similarly to the bare Dirac cone, the Born contribution decays as $E_F^{-1}$ as the Fermi energy moves away from the DP. In the parameter range shown, for $E_F\gtrsim30$ meV the particle-hole-asymmetry corrections and skew scattering therefore dominate the remaining orbital current.

\section{Discussion}
There are two central points. First, the wavevector-off-diagonal kinetic equation provides a systematic route to disorder corrections in the full internal orbital-current expectation value. Second, disorder-generated terms are quantitatively important. Contributions of order $\tau^0$ have the same disorder scaling as the intrinsic response, so varying the disorder strength alone cannot distinguish them. When particle-hole symmetry and the disorder ensemble allow conventional skew scattering, the skew term is of order $\tau$, with $j_{\rm skew}\propto U_0\tau$, and therefore grows as $1/n_i$ at fixed impurity potential. It can consequently dominate in the clean limit. For the models studied here the field-corrected side-jump contribution is small. The resulting non-crossing hierarchy therefore mirrors the familiar anomalous- and spin-Hall problems: intrinsic and disorder-generated $\tau^0$ terms compete, while the $\tau$-linear skew contribution can dominate sufficiently clean samples. The exact cancellation in the bare Dirac cone is model specific, but this competition and the wavevector-off-diagonal construction are not. Crossing diagrams can modify the complete $\tau^0$ coefficient, but not this scaling distinction or the need for the wavevector-off-diagonal density matrix.

Fig.~\ref{MDC_OHE} shows that in the bare massive Dirac cone model the Born and intrinsic contributions to the orbital current peak near the Dirac point (DP) and decrease $\propto E_F^{-1}$, whereas Fig.~\ref{Total_ext_TMDCPH} shows the skew-scattering contribution and the quadratic corrections increase as the Fermi energy moves away from the DP. The massive Dirac cone model describes the band structure of a number of materials of interest, such as the surface states of magnetic topological insulators and transition metal dichalcogenides. Experimentally, this indicates that in clean samples a large OHE can be expected at all Fermi energies in the vicinity of the DP, whereas in dirtier samples the OHE will be sharply peaked in the vicinity of the DP. The calculations in this paper use effective $\bm k\cdot\bm p$ models because the kinetic-equation formalism can be applied to them directly. However, recent work has shown that while such effective models capture the band structure accurately, they often cannot completely capture the orbital magnetisation \cite{lee2026anatomy}. By extension, an effective model need not capture every contribution to the OHE quantitatively. Extending the present disorder formalism to realistic first-principles Hamiltonians is therefore an important next step. The model calculations nevertheless establish the main qualitative point: impurity corrections can be comparable to the intrinsic current, and when skew scattering is allowed it grows relative to the $\tau^0$ terms in the clean limit.

A major motivation for the OHE is the generation of orbital torque. Such a torque requires the injected OAM to couple ultimately to the magnetisation, commonly through orbital-to-spin conversion. The microscopic conversion mechanisms remain incompletely understood for both itinerant and local OAM. On-site spin-orbit coupling of the form $\alpha\bm L\cdot\bm s$ provides a natural local conversion channel, while the transmission and conservation of Bloch OAM across interfaces remain open problems. Orbital Hall conductivities should therefore be regarded as indicators of the available bulk OAM flux rather than direct quantitative predictions of torque efficiency. 

Extrinsic scattering corrections to the orbital current in the massive Dirac cone model have been studied previously in Refs.~\cite{Hong-OHE-PRL,veneri2024extrinsic}, however, these calculations use the conventional approach to the orbital current. The conventional approach includes only the OAM matrix elements that appear in the equilibrium OAM expectation value. This approach neglects matrix elements that have since been shown to be important \cite{liu2025quantumOHE}. For this model the conventional approach is equivalent to retaining only the band-off-diagonal position matrix elements, $r^{i,mn}_{\bm k}=\mathcal{R}^{i,mn}_{\bm k}$ for $m\neq n$. The corresponding current can be written as 
\begin{equation}\label{Eq:OldCalc}
    \langle j^a_\delta\rangle_{\hat{r}\text{ off-diagonal}}=\frac{\epsilon_{abc}}{4}\text{Tr}\!\left[ \{v_c, v_\delta\}\{\mathcal{R}_b, \rho\} + [v_c ,[\mathcal{R}_b,v_\delta]]\rho\right] .
\end{equation}
When comparing the above expression with (\ref{Eq:OHcurrent}) it is clear that the commutator in (\ref{Eq:OldCalc}) is absorbed into the covariant derivative of velocity. This is the primary source of the difference between these results and the results of Ref.~\cite{Hong-OHE-PRL}, as the massive Dirac cone model is linear in wavevector, so the covariant derivative of velocity is automatically zero, whereas the commutator of $v$ with $\mathcal{R}$ is nonzero. This difference further demonstrates the importance of evaluating the full orbital-current operator rather than truncating the position operator to its interband matrix elements.

Evaluating disorder corrections to the full orbital current is technically more demanding than the conventional calculation because expectation values involving the position operator require the wavevector-off-diagonal density matrix \cite{liu2025quantumOHE, Hong-PSHE-PRB, PhysRevB.108.245418}. After disorder averaging, translational symmetry is restored in the impurity correlators, while the infinitesimal wavevector mismatch is carried by $\rho_{\bm k_+\bm k_-}$. Our procedure is therefore to derive each disorder-averaged collision term at finite $\bm Q$, expand it to linear order, solve the resulting hierarchy for the $\bm Q^0$ and $\bm Q^1$ density matrices, and finally insert them into Eqs.~(\ref{Eq:OHcurrent}) and (\ref{Eq:Xi}). The full expressions are given in the Supplemental Material.

A notable feature of the full calculation is the appearance, before the internal-OAM projection is made, of free-streaming contributions that scale as $\tau^2$ and $\tau$. The $\tau^2$ piece combines the Drude occupation shift $n^{(1)}_{\bm E,0}\propto\tau$ with a further free-flight displacement accumulated over a collision time; the analogous $\tau$ contribution is generated when the already existing $n^{(0)}_{\bm E,0}$ streams between collisions. These terms are legitimate parts of the wavevector-off-diagonal density matrix and are retained in the kinetic equation, where they can feed later scattering processes. Their contribution is $\bm\Xi^{\rm CM}$ and is absent from the lever arm $\bm\Xi^{\rm int}$ by the definition of the internal-OAM current. This removes the origin-dependent centre-of-mass contribution $\Delta\bm R_{\rm CM}\times\bm v$ without discarding the local electric-field-induced dipole or other diagonal pieces of $\bm\Xi$. The distinction is specific to the observable, not to the kinetic equation. It also highlights a broader issue shared with spin transport: the microscopic current, a conserved current, and the boundary accumulation are distinct objects, and the relation among them depends on the geometry and on angular-momentum non-conservation. Here our aim is narrower: to evaluate the disorder corrections to the full internal orbital-current operator after separating its centre-of-mass free-flight lever arm. Finally, OAM, like spin, is not separately conserved and, in analogy with the proper definition of the spin current \cite{PhysRevB.108.245418, Sinova-Dimi-PRL-2004, Defintion-SC-PRL-2006-Qian}, a complete understanding of OAM transport necessarily requires one to evaluate the conserved current operator that incorporates the corresponding torque dipole contribution, with all its attendant complications. Such a description remains outside the scope of the present work.

\section{Conclusion}
We have developed a microscopic theory of extrinsic contributions to the full internal orbital current, retaining all matrix elements of the position operator and solving the quantum kinetic equation for the wavevector-off-diagonal density matrix. Within the non-crossing approximation the formalism treats Born, field-corrected side-jump, and skew-scattering terms on the same footing. The complete density matrix is retained throughout; the explicitly identifiable free-flight centre-of-mass pieces form $\bm\Xi^{\rm CM}$ and are absent from the lever arm $\bm\Xi^{\rm int}$ by the definition of the internal-OAM observable. In the bare massive Dirac cone the surviving non-crossing extrinsic correction gives $j_L=2 \Delta j_{1}$, while particle-hole asymmetry activates conventional skew scattering for a disorder ensemble with a non-zero third moment. More generally, disorder-generated $\tau^0$ contributions can compete directly with the intrinsic response and cannot be separated from it by simple disorder-strength scaling. The projector formulation extends the construction to degenerate multiband systems, so quantitative predictions of the orbital Hall effect require disorder and the full position operator to be treated consistently.

\funding{This work is supported by the Australian Research Council Discovery Project DP2401062.}

\roles{J.~H.~C performed the orbital Hall effect calculations. J.~H.~C made the figures. J.~H.~C and D.~C wrote the manuscript. D.~C supervised the project.}

\data{The authors declare that the data supporting the findings of this study are available within the paper and its supplementary information file.}

\suppdata{Supplementary material accompanies this article.}

\bibliographystyle{iopart-num}
\bibliography{OAM}

\end{document}